\documentclass[12pt]{article}

\usepackage{amsmath,amssymb,amsfonts,fancyhdr,fancybox,graphics,epsfig,calc,color,subfigure}

\usepackage{amsbsy}
\usepackage{mathrsfs}

\usepackage{chngpage}

\usepackage{lipsum}

\usepackage{booktabs}

\usepackage{multicol}

\usepackage{caption}

\renewcommand{\baselinestretch}{1.4}
\def\blist#1#2#3
{    \begin{list}{}{
     \setlength{\parsep}{0pt}
     \setlength{\leftmargin}{1in}
     \setlength{\listparindent}{0pt}
     \setlength{\itemindent}{\listparindent}
     \setlength{\labelsep}{#3 pt}
     \setlength{\labelwidth}{\leftmargin}
     \addtolength{\labelwidth}{-\labelsep}
     \addtolength{\labelwidth}{\itemindent}

     \setlength{\rightmargin}{1in}}   }

\def\elist{\end{list}}

\newcommand{\argmin}{\operatornamewithlimits{argmin}}

\usepackage{amsmath}
\def\boxit#1{\vbox{\hrule\hbox{\vrule\kern6pt\vbox{\kern6pt#1\kern6pt}\kern6pt\vrule}\hrule}}

\begin{document}

\topmargin -0.3in \oddsidemargin 0.2in \evensidemargin 0.2in
\baselineskip 9mm
\renewcommand \baselinestretch {1.2}

\title{Tuning Parameter Selection in Regularized Estimations of Large Covariance Matrices}

\author{Yixin Fang$^{\ast 1}$, Binhuan Wang$^1$, and Yang Feng$^2$\\
{\it $^1$New York University and $^2$Columbia University}}

\footnotetext[1]{
Address for Correspondence: Division of Biostatistics, NYU School of Medicine, 650 First Avenue, 5th floor, New York, NY 10016. Email: yixin.fang@nyumc.org.}

\date{(\today)}
\maketitle

\begin{abstract}

Recently many regularized estimators of large covariance matrices have been proposed, and the tuning parameters in these estimators are usually selected via cross-validation. However, there is no guideline on the number of folds for conducting cross-validation and there is no comparison between cross-validation and the methods based on bootstrap. Through extensive simulations, we suggest 10-fold cross-validation (nine-tenths for training and one-tenth for validation) be appropriate when the estimation accuracy is measured in the Frobenius norm, while 2-fold cross-validation (half for training and half for validation) or reverse 3-fold cross-validation (one-third for training and two-thirds for validation) be appropriate in the operator norm. We also suggest the ``optimal" cross-validation be more appropriate than the methods based on bootstrap for both types of norm.

\end{abstract}

\bigskip

{\bf Keywords:} Banding; Bootstrap; Covariance matrix; Cross-validation; Frobenius norm; Operator norm; Thresholding.

\newpage
\setcounter{equation}{0}

\section{Introduction}

Estimation of covariance matrices is important in many statistical areas including principal component analysis, linear discriminant analysis, and graphical modeling. Recently, these tools have been used for analyzing high-dimensional datasets where the sample sizes can be much smaller than the dimensions. Examples include image data, genetic data, and financial data.

Suppose that there are $n$ identically distributed $p$-dimensional random variables $X_1, \ldots, X_n$ with covariance matrix $\Sigma_{p\times p}$. It is well known that the empirical covariance matrix,
\begin{equation}
\widetilde{\Sigma}=[\tilde{\sigma}_{ij}]=\frac{1}{n}\sum_{i=1}^{n}(X_i-\overline{X})(X_i-\overline{X})^{\top}, \label{Esitmate_0}
\end{equation}
where $\overline{X}=\sum_{i=1}^n X_i/n$, is not a good estimator of $\Sigma$ when $p>n$. To overcome the curse of dimensionality, many regularized estimators of large covariance matrices have been proposed recently; see Johnstone (2001) and references therein.

\subsection{Two groups of estimators}

There are two main groups of such estimators. One group assumes the covariance matrix being estimated is sparse in the sense that many entries are zero or nearly so. Methods in this group include thresholding (Bickel and Levina, 2008a; El Karoui, 2008) and generalized thresholding (Rothman, Levina, and Zhu, 2009).

Bickel and Levina (2008a) studied the asymptotic properties of the hard-thresholding estimator,
\begin{equation}
\widehat{\Sigma}^{HT}(\lambda)=[\tilde{\sigma}_{ij}I(|\tilde{\sigma}_{ij}|\geq \lambda)],\label{hard-threshold}
\end{equation}
where $\lambda>0$ is a tuning parameter to be selected. For the ease of presentation, we use the same notation $\lambda$ for the tuning parameters in all different kinds of estimators. Rothman {\it et al.}~(2009) proposed a class of thresholding estimators, including the soft-thresholding estimator,
\begin{equation}
\widehat{\Sigma}^{ST}(\lambda)=[\mbox{sign}(\tilde{\sigma}_{ij})(|\tilde{\sigma}_{ij}|- \lambda)_+].\label{soft-threshold}
\end{equation}

The other group is for applications where there is a natural metric on the dimensional index set and one expects that the entries farther away from diagonal are smaller. Methods in this group include banding (Bickel and Levina, 2008b; Wu and Pourahmadi, 2009) and tapering (Furrer and Bengtsson, 2007; Cai, Zhang, and Zhou, 2010).

Bickel and Levina (2008b) studied the asymptotic properties of the banding estimator,
\begin{equation}
\widehat{\Sigma}^{Ba}(\lambda)=[\tilde{\sigma}_{ij}I(|i-j|\leq \lambda)],\label{banding}
\end{equation}
where integer $0\leq \lambda <p$ is a tuning parameter. Cai {\it et al.} (2010) studied the asymptotic properties of the tapering estimator,
\begin{equation}
\widehat{\Sigma}^{Ta}(\lambda)=[w_{ij}^{\lambda}\tilde{\sigma}_{ij}],\label{tapering}
\end{equation}
where, for integer $0\leq \lambda <p$, $w_{ij}^{\lambda}=1$ when $|i-j|\leq \lambda/2$, $w_{ij}^{\lambda}=2-2|i-j|/\lambda$ when $\lambda/2<|i-j|<\lambda$, and $w_{ij}^{\lambda}=0$ otherwise.

In this work, we focus on these four estimators, although there are many other methods not belonging to these two groups, such as Cholesky-based regularization (Huang {\it et al.}, 2006; Lam and Fan, 2009; Rothman, Levina, and Zhu, 2010) and factor-based regularization (Fan, Fan, and Lv, 2008; Fan, Liao, Mincheva, 2013).

\subsection{Tuning parameter selection}

The performance of any estimator depends  heavily  on the quality of tuning parameter selection. There are two popular norms which can be used to measure the estimation accuracy, one is the Frobenius norm and the other is the operator norm. For any matrix $M_{p\times p}=[m_{ij}]$, its Frobenius norm and operator norm are defined as
\begin{equation}
\| M \|_F=\sqrt{\sum_{i=1}^p\sum_{j=1}^pm_{ij}^2} \mbox{\ \ and \ } \|M\|_{op}=\sup\{\|M{\bf x}\|_2: \|{\bf x}\|_2=1\},\label{norms}
\end{equation}
respectively, where $\|\cdot\|_2$ is the Euclidean norm for vectors.

If the estimand $\Sigma$ were known, for any estimator $\widehat{\Sigma}(\lambda)$, we would select $\lambda$ as
\begin{equation}
\lambda^{{oracle}}_{F}=\underset{\lambda}{\argmin}\|\widehat{\Sigma}(\lambda)-\Sigma\|_F \mbox{\ \ or \ } \lambda^{{oracle}}_{op}=\underset{\lambda}{\argmin}\|\widehat{\Sigma}(\lambda)-\Sigma\|_{op}, \label{oracle}
\end{equation}
depending on which norm is considered. In practice, we attempt to estimate the Frobenius risk or the operator risk first,
\begin{equation}
R_{F}(\lambda)=\mathbb{E}\|\widehat{\Sigma}(\lambda)-\Sigma\|^2_F \mbox{\ \ or \ } R_{op}(\lambda)=\mathbb{E}\|\widehat{\Sigma}(\lambda)-\Sigma\|^2_{op}, \label{oracle}
\end{equation}
and then select a value for $\lambda$.

The remaining of the manuscript is organized as follows. In Section 2, we describe two popular methods, cross-validation and bootstrap, for estimating the risk under consideration. In Section 3, we conduct extensive simulations to provide some evidences about how many folds for cross-validation would be ``optimal" and whether the ``optimal" cross-validation might be better than the methods based on bootstrap. Some conclusions are summarized in Section 4 and the Appendix contains all the technical proofs.

\section{Methods}

\subsection{Cross-validation}

Since 1970s (e.g., Stone, 1974), cross-validation has become one of the most popular methods for tuning parameter selection. Especially for regularized estimators of large covariance matrices, cross-validation plays a dominant role in tuning parameter selection. $V$-fold cross-validation first splits data into $\{\mathcal{D}_1, \ldots, \mathcal{D}_V\}$, and then selects the tuning parameter in $\widehat{\Sigma}(\lambda)$ as
\begin{equation}
\lambda^{{cv}}_F=\underset{\lambda}{\argmin}\frac{1}{V}\sum_{v=1}^V\|\widehat{\Sigma}^{(-v)}(\lambda)-\widetilde{\Sigma}^{(v)}\|^2_F \mbox{\ \ or \ } \lambda^{{cv}}_{op}=\underset{\lambda}{\argmin}\frac{1}{V}\sum_{v=1}^V\|\widehat{\Sigma}^{(-v)}(\lambda)-\widetilde{\Sigma}^{(v)}\|^2_{op}, \label{CV}
\end{equation}
where $\widetilde{\Sigma}^{(v)}$ is the un-regularized estimator (\ref{Esitmate_0}) based on $\mathcal{D}_v$ and $\widehat{\Sigma}^{(-v)}(\lambda)$ is the regularized estimator under consideration based on data without $\mathcal{D}_v$.  Here the size of training data is about $(V-1)n/V$ and the size of validation data is about $n/V$.

Shao (1993) argued that for linear models the cross-validation is asymptotically consistent if the ratio of validation size over sample size goes to one. Motivated by this result, we also consider reverse cross-validation to select the tuning parameter in $\widehat{\Sigma}(\lambda)$ as
\begin{equation}
\lambda^{{rcv}}_F=\underset{\lambda}{\argmin}\frac{1}{V}\sum_{v=1}^V\|\widehat{\Sigma}^{(v)}(\lambda)-\widetilde{\Sigma}^{(-v)}\|^2_F \mbox{\ \ or \ } \lambda^{{rcv}}_{op}=\underset{\lambda}{\argmin}\frac{1}{V}\sum_{v=1}^V\|\widehat{\Sigma}^{(v)}(\lambda)-\widetilde{\Sigma}^{(-v)}\|^2_{op}, \label{RCV}
\end{equation}
where $\widetilde{\Sigma}^{(-v)}$ is the un-regularized estimator (\ref{Esitmate_0}) based on data without $\mathcal{D}_v$ and $\widehat{\Sigma}^{(v)}(\lambda)$ is the regularized estimator under consideration based on $\mathcal{D}_v$.  Here the size of training data is about $n/V$ and the size of validation data is about $(V-1)n/V$.

However, there is a lack of consensus, even discussion, on how many folds should be considered when using cross-validation (or reverse cross-validation) to select tuning parameters in the regularized estimators of large covariance matrices. Here are some examples. In Bickel and Levina (2008a), 2-fold cross-validation was used (i.e., the training size is $n_1=n/2$ and the validation size is $n_2=n/2$). In Bickel and Leveina (2008b), reverse 3-fold cross-validation was used (i.e., $n_1=n/3$ and $n_2=2n/3$). In Yi and Zou (2012) and Xue, Ma, and Zou (2013), 5-fold cross-validation was used (i.e., $n_1=4n/5$ and $n_2=n/5$).

\subsection{Bootstrap}

\subsubsection{Bootstrap for the Frobenius norm}

Let $\widetilde{\Sigma}^{s}=\frac{n}{n-1}\widetilde{\Sigma}=[\tilde\sigma_{ij}^s]$ be the usual sample covariance matrix. Let $\widehat{\Sigma}(\lambda)=[\hat{\sigma}_{ij}(\lambda)]$ be the regularized estimator under consideration. In their Lemma 1, Yi and Zou (2012) showed that the Frobenius risk can be decomposed into
\begin{eqnarray}
R_F(\lambda)&=&\mathbb{E}\|\widehat{\Sigma}(\lambda)-\widetilde{\Sigma}^{s}\|^2_F+2\sum_{i=1}^p\sum_{j=1}^p\mbox{Cov}(\hat{\sigma}_{ij}(\lambda), \tilde\sigma_{ij}^s)-\sum_{i=1}^p\sum_{j=1}^p\mbox{var}(\tilde\sigma_{ij}^s),\label{decomp}\\
&=&\mbox{apparent\ error} \ \hskip 1pt + \ \ \ \ \mbox{covariance\ penalty} \ \ \ \ \hskip -1pt - \ \ \ \ \mbox{constant}\nonumber
\end{eqnarray}
where terms ``apparent error" and ``covariance penalty" come from Efron (2004). In the same paper, Efron proposed to use the bootstrap method to estimate the covariance penalty. Assume that $\{X_1^{b*}, \ldots, X_n^{b*}\}$, $b=1, \ldots, B$,
are samples repeatedly drawn from some underlying parametric or non-parametric bootstrap model (to be discussed later). For each bootstrap sample, the corresponding estimates $\widetilde{\Sigma}^{s, b*}=[\tilde{\sigma}_{ij}^{s, b*}]$ and $\widehat{\Sigma}(\lambda)^{b*}$ are obtained. Then the covariance penalty can be estimated by
\begin{equation}
\widehat{\mbox{Cov}}(\lambda)=2\sum_{i=1}^p\sum_{j=1}^p\Big(\frac{1}{B-1}\sum_{b=1}^B\hat{\sigma}^{b*}_{ij}(\lambda)
\tilde\sigma_{ij}^{s, b*}-\frac{1}{B(B-1)}\sum_{b=1}^B\hat{\sigma}^{b*}_{ij}(\lambda) \sum_{b=1}^B\tilde\sigma_{ij}^{s, b*}\Big),\label{cov_est}
\end{equation}
and the Frobenius risk can be estimated by
\begin{equation}
\widehat{R}_F(\lambda)=\|\widehat{\Sigma}(\lambda)-\widetilde{\Sigma}^{s}\|^2_F+\widehat{\mbox{Cov}}(\lambda),\label{risk_est}
\end{equation}
where the constant term in (\ref{decomp}) can be ignored for tuning parameter selection and can be recovered for risk estimation. Then the tuning parameter can be selected as
\begin{equation}
\lambda^{{boot}}_F=\underset{\lambda}\argmin \widehat{R}_F(\lambda).\label{F_boot}
\end{equation}

Now let's discuss how to select an appropriate bootstrap model for generating bootstrap samples. First, for high-dimensional applications, parametric bootstrap is better than non-parametric bootstrap. Second, as pointed out by Efron (2004), the ``ultimate bigger" bootstrap model,
\begin{equation}
\widehat{\bf F}=N\big(\overline{X}, \widetilde{\Sigma}^s\big), \label{ultimate}
\end{equation}
where $N$ stands for multi-normal distribution, has ``the advantage of not requiring model assumptions", but ``pays for this luxury with increased estimation error". Third, as pointed out also by Efron (2004), ``the exact choice of $\widehat{\bf F}$ is often quite unimportant". Considering these remarks, in all the numerical results, we consider an intermediate bootstrap model,
\begin{equation}
\widehat{\bf F}=N\big(\overline{X}, \widehat{\Sigma}(\hat\lambda_0)\big), \label{inter_boot}
\end{equation}
where $\hat\lambda_0$ is selected via (\ref{F_boot}) based on the ultimate bootstrap model (\ref{ultimate}).

It should be pointed out that for banding estimator (\ref{banding}) and tapering estimator (\ref{tapering}), Yi and Zou (2012) derived an explicit formula for the covariance penalty in (\ref{decomp}) and proposed an unbiased estimator for it. They called this method ``SURE-tuned" estimation, and is similar to bootstrap using the ultimate bootstrap model (\ref{ultimate}).

\subsubsection{Bootstrap for the operator norm}

It is very difficult to estimate the operator risk $R_{op}(\lambda)$, because it cannot be easily decomposed like (\ref{decomp}). Here we derive a rough approximation to $R_{op}(\lambda)$ for banding and tapering estimators and hope this will stimulate more accurate approximations.

For any regularized estimator $\widehat{\Sigma}(\lambda)$, let $\Gamma=(\widehat{\Sigma}(\lambda)-\Sigma)(\widehat{\Sigma}(\lambda)-\Sigma)^{\top}$ and $\Gamma^*=\mathbb{E}(\Gamma)$. Following the delta-method in Silverman (1996) and some arguments in Appendix, we have
\begin{equation}
R_{op}(\lambda)=\mathbb{E}\big(\underset{\|\beta\|_2=1}{\max}\beta^{\top}\Gamma\beta\big)\doteq\underset{\|\beta\|_2=1}
{\max}\beta^{\top}\mathbb{E}(\Gamma)\beta+\beta_1^{*\top} \mathbb{E}\big(\Delta\Pi\Delta\big)\beta_1^*,\label{approx}
\end{equation}
where $\Delta=\Gamma-\Gamma^*$ and $\Pi=\sum_{j=2}^p\frac{1}{l^*_1-l_j^*}\beta_j^*\beta_j^{*\top}$ with $\{(\beta_j^*, l_j^*), j=1, \ldots, p\}$ being the eigenvectors and eigenvalues from eigen-system $\Gamma^*\beta=l\beta$. The last term in (\ref{approx}) is known as {\it Hadamard second variation formula} (e.g., Tao, 2012). The approximation still holds if $\Gamma^*$ is replaced by some unbiased estimator $\widehat{\Gamma}^*$; i.e.,
\begin{eqnarray}
R_{op}(\lambda)&\doteq&\hat{l}_1^*+\hat{\beta}_1^{*\top} \mathbb{E}\big(\widehat{\Delta}\widehat{\Pi}\widehat{\Delta}\big)\hat{\beta}_1^*,\label{approx_est}\\
&\doteq& \mbox{apparent\ error}+\mbox{covariance\ penalty.}\nonumber
\end{eqnarray}
where $\widehat{\Delta}=\Gamma-\widehat{\Gamma}^*$ and $\widehat{\Pi}=\sum_{j=2}^p(\hat{l}^*_1-\hat{l}_j^*)^{-1}\hat\beta_j^*\hat\beta_j^{*\top}$ with $\{(\hat\beta_j^*, \hat{l}_j^*), j=1, \ldots, p\}$ from eigen-system $\widehat{\Gamma}^*\beta=l\beta$. Furthermore, we can estimate the expectation in the second term of (\ref{approx_est}) via the bootstrap using the same model as (\ref{inter_boot}).

{\it Remark 1:} For banding estimator (\ref{banding}) and tapering estimator (\ref{tapering}), we derive an unbiased estimator $\widehat{\Gamma}^*$ in Appendix. Unfortunately, we fail to derive any unbiased estimator for thresholding estimators (\ref{hard-threshold}) and (\ref{soft-threshold}).  {\it Remark 2:} It may be inappropriate to call the two terms in (\ref{approx_est}) ``apparent error" and ``covariance penalty", but it is helpful for understanding the bias-variance trade-off in $R_{op}(\lambda)$. {\it Remark 3:} Based on our limited numerical expericence, the approximation in (\ref{approx}) is very accurate, but due to curse of dimensionality, the approximation in (\ref{approx_est}) is rough for high-dimensional data.

\section{Simulation results}

The data are generated from ${N}(0, \Sigma)$ with three covariance models adopted from Yi and Zou (2012) are considered, and four settings of sample size and dimension are considered; that is, $(n, p)=(100, 100), (100, 200), (200, 100)$ and $(200, 200)$.

{\it Model 1.} The covariance matrix $\Sigma=[\sigma_{ij}]$, where $\sigma_{ii}=1$ for $1\leq i\leq p$ and $\sigma_{ij}=\rho|i-j|^{-(\alpha+1)}$ for $1\leq i\neq j\leq p$. Let $\rho=0.6$ and $\alpha=0.1$ or $0.5$.

{\it Model 2.} The covariance matrix $\Sigma=[\sigma_{ij}]$, where $\sigma_{ij}=\rho^{|i-j|}$ for any $1\leq i, j\leq p$. Let $\rho=0.9$ or $0.5$.

{\it Model 3.} This model is a truncated version of model 1, where $\sigma_{ii}=1$ for $1\leq i\leq p$ and $\sigma_{ij}=\rho|i-j|^{-(\alpha+1)}I(|i-j|\leq 6)$ for $1\leq i\neq j\leq p$. Let $\rho=0.6$ and $\alpha=0.1$ or $0.5$.

Eight cross-validation methods for tuning parameter selection are compared: 2-fold, 3-fold, 5-fold, and 10-fold cross-validations (CV2, CV3, CV5, CV10), 2-fold cross-validation based on 50 random splits (RCV2), reverse 3-fold, 5-fold, and 10-fold cross-validations (reCV3, reCV5, reCV10), along with the bootstrap methods (bootstrap). The cross-validation using $n_1=n-n/\log(n)$ for training and $n_2=n/\log(n)$ for validation is also compared, but the results are not reported because this method does not perform very well compared with others although some nice asymptotic property was derived in Bickel and Levina (2008a).

We use the Frobenius norm and the operator norm as evaluation criteria with all four regularized estimators in Section 1 considered. Each simulation setting is repeated $K=200$ times, and the performance is measured by the empirical Mean Square Error (MSE), which is the average of $200$ values of $\|\widehat{\Sigma}(\hat{\lambda})-\Sigma\|^2_{F}$ or  $\|\widehat{\Sigma}(\hat{\lambda})-\Sigma\|^2_{op}$.

\subsection{Results in the Frobenius norm}

First, eight cross-validation methods are compared using the Frobenius norm, with results summarized in Figures 1-3. Only the results from the setting where $(n, p)=(200, 200)$ are reported here; results from the other three settings are similar and can be found in the Supplement. If some MSE values are too big, they are excluded from the figures. Since all the true covariance matrices have some banding or tapering structure, both the banding estimator and the tapering estimator are more accurate than the thresholding estimators.

From Figures 1-3, we see that 10-fold cross-validation performs best for all three models and all four regularized estimators. Also we see that 5-fold cross-validation performs comparably with 10-fold cross-validation, so maybe it is unnecessary to consider folds more than 10. This finding is quite similar to the one in Kohavi (1995), which also suggested 10-fold cross-validation for linear models.

Then 10-fold cross-validation method is compared with the bootstrap method in Subsection 2.2.1 and the SURE method in Yi and Zou (2012), with results summarized in Figures 4-6. In Yi and Zou (2012), the SURE method was compared with 5-fold cross-validation and it was found that the SURE method performs slightly better than 5-fold cross-validation. However, from Figures 4-6, we see that 10-fold cross-validation performs slightly better than the SURE method. This is not a conflict because from Figures 1-3 we just see that 10-fold cross-validation performs slightly better that 5-fold cross-validation. Also note that the SURE method is only applicable for the banding and tapering estimators.

From Figures 4-6, we also see that the bootstrap method performs very similar to 10-fold cross-validation for the banding and tapering estimators. However, the comparison is complicated for thresholding estimators, because sometimes 10-fold cross-validation performs much better than the bootstrap method whereas sometimes the bootstrap method performs slightly better than 10-fold cross-validation.

\subsection{Results in the operator norm}

Again, eight cross-validation methods are compared using the operator norm, with results summarized in Figures 7-9. Similarly, since all the true covariance matrices have some banding or tapering structure, both the banding estimator and the tapering estimator are more accurate than the thresholding estimators.

For Figure 7-9, we see that, for the banding and tapering estimators, reverse 3-fold cross-validation performs best in most cases, while in other cases it performs almost as well as reverse 5-fold cross-validation. For the hard-thresholding estimator, 2-fold cross-validation or 2-fold cross-validation based on 50 random splits performs the best in all the cases. For the soft-thresholding estimator, either 2-fold cross-validation or reverse 3-fold cross-validation performs best in all the cases. In addition, it seems using multiple random splits does not improve the performance significantly.

Therefore, from Figure 7-9, we see that either 2-fold cross-validation or reverse 3-fold cross-validation performs best if the MSE is in terms of the operator norm. This finding is different from the result that 10-fold cross-validation performs best for the Frobenius norm. In other words, we need bigger training size for the Frobenius from whereas we need bigger validation size for the operator norm.

Then, for the banding and the tapering estimators, reverse 3-fold cross-validation method is compared with the bootstrap method in Subsection 2.2.2, with results summarized in Figures 10-12. We see that the bootstrap methods performs comparably well as reverse 3-fold cross-validation for the banding and tapering estimators. The comparison shows that the rough approximation (\ref{approx_est}) is working well. On the other hand, since the bootstrap does not outperform reverse 3-fold cross-validation and it is much more computationally expensive, we recommend the reverse 3-fold cross-validation over the bootstrap method when the operator norm is considered.

\section{Conclusions}

In this manuscript, we compare two methods (cross-validation and bootstrap) for selecting tuning parameters in two groups of regularized estimators (banding and thresholding) for covariance matrix, where estimation accuracy is measured in two norms (Frobenius norm and operator norm). Based on extensive simulations, we draw the following conclusions:

1. Cross-validation is computational convenient and performs better than (or as well as) the methods based on bootstrap.

2. If the Frobenius norm is considered, we suggest 10-fold cross-validation for both groups of regularized estimators.

3. If the operator norm is considered, we suggest 2-fold cross-validation for the thresholding estimators and reverse 3-fold cross-validation for the banding and tapering estimators.

\begin{center}
{\bf\Large {Appendix}}
\end{center}
\renewcommand{\theequation}{A.\arabic{equation}}
\setcounter{equation}{0}

\noindent{\bf A.1 Approximation in (\ref{approx})}

Let $\{(\beta_j, l_j), j=1, \ldots, p\}$ be eigenvectors and eigenvalues from eigen-system $\Gamma\beta=l\beta$. Following the delta method used in Silverman (1996), let
\begin{eqnarray*}
\Gamma&=&\Gamma^*+\varepsilon \Delta^0,\\
\beta_1&=&\beta_1^*+\varepsilon\beta^{*(1)}+\alpha\beta^{*(2)}+\ldots,\\
l_1&=&l^*_1+\varepsilon l^{*(1)}+\alpha l^{*(2)}+\ldots,
\end{eqnarray*}
where $\varepsilon=n^{-1/2}$ and $\alpha=n^{-1}$. Because $\Gamma\beta_1=l_1\beta$, we have
\begin{eqnarray*}
(\Gamma^*+\varepsilon \Delta^0)(\beta_1^*+\varepsilon\beta^{*(1)}+\alpha\beta^{*(2)}+\ldots)=(l^*_1+\varepsilon l^{*(1)}+\alpha l^{*(2)}+\ldots)(\beta_1^*+\varepsilon\beta^{*(1)}+\alpha\beta^{*(2)}+\ldots).
\end{eqnarray*}
Comparing the coefficients of powers of $\varepsilon$ and $\alpha$ on both sides of this equation, we have
\begin{eqnarray*}
l_1\doteq l_1^*+\varepsilon \beta_1^{*\top}\Delta^0\beta_1^*+\alpha \beta_1^{*\top}\Delta^0\Pi\Delta^0\beta_1^*,
\end{eqnarray*}
whose expectation is (\ref{approx}).

\noindent{\bf A.2 Unbiased estimator of $\Gamma$}

With a slight abuse of notation, for both the banding estimator and the tapering estimator, let $$\widehat{\Sigma}(\lambda)=[w_{ij}^{\lambda}\tilde{\sigma}_{ij}^{s}].$$
Let $W^{\lambda}=[w_{ij}^{\lambda}]$, whose $j$th column is defined as $w_j^{\lambda}$. Also let $\Sigma_j$ be the $j$th column of $\Sigma$ and let  $W^{\lambda}_{j}=\mbox{diag}(w_j^{\lambda})$. Note that $(n-1)\widetilde{\Sigma}^s\sim W_p(n-1, \Sigma)$, where $W$ stands for Wishart distribution. By some tedious arguments, we have
$$\Gamma^*=\frac{1}{n-1}\sum_{j=1}^pW_{j}^{\lambda}[\sigma_{jj}\Sigma+\Sigma_j\Sigma_j^{\top}]W_{j}^{\lambda}+
\sum_{j=1}^p(W_{j}^{\lambda}-I_p)\Sigma_j\Sigma_j^{\top}(W_{j}^{\lambda}-I_p).$$
In order to find an unbiased estimator for $\Sigma^*$, it suffices to find an unbiased estimator for $\sigma_{kl}\sigma_{k'l'}$ for any $1\leq k, l, k', l'\leq p$. Let $X_i=(X_{i1}, \ldots, X_{ip})^{\top}$, $\overline{X}_{j}=\sum_{i} X_{ij}/n$, and $\overline{X}^{(-i)}_{j}=\sum_{i'\neq i} X_{i'j}/(n-1)$. In this manuscript, we use
$$\frac{1}{n-1}\sum_{i=1}^n(X_{ik}-\overline{X}_k)(X_{il}-\overline{X}_l)\times \frac{1}{n-2}\sum_{i'\neq i}(X_{i'k'}-\overline{X}^{(-i)}_{k'})(X_{i'l'}-\overline{X}^{(-i)}_{l'})$$
to estimate $\sigma_{kl}\sigma_{k'l'}$.



\newpage
\begin{figure}[h]\caption{\small Eight cross-validation methods are compared for Model 1 ($n=200$, $p=200$, $\alpha=0.1$ in left panel and $\alpha=0.5$ in right panel). Performances are measured by MSE in Frobenius norm.}\label{F-M1-CV}
\begin{center}
\includegraphics[width=6in, height=4in]{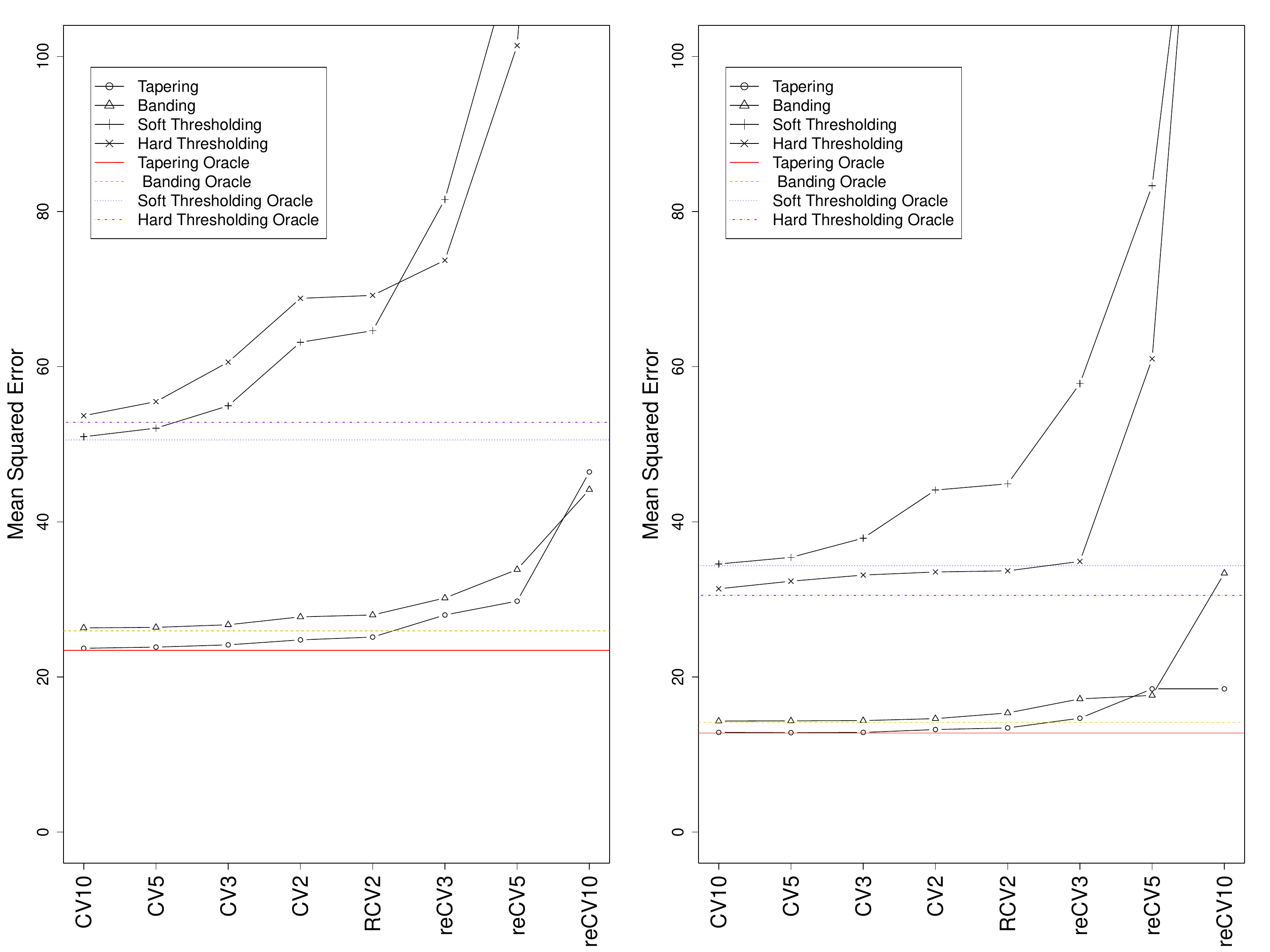}
\end{center}
\end{figure}

\newpage
\begin{figure}[h]\caption{\small Eight cross-validation methods are compared for Model 2 ($n=200$, $p=200$, $\rho=0.9$ in left panel and $\rho=0.5$ in right panel). Performances are measured by MSE in Frobenius norm.}\label{F-M2-CV}
\begin{center}
\includegraphics[width=6in, height=4in]{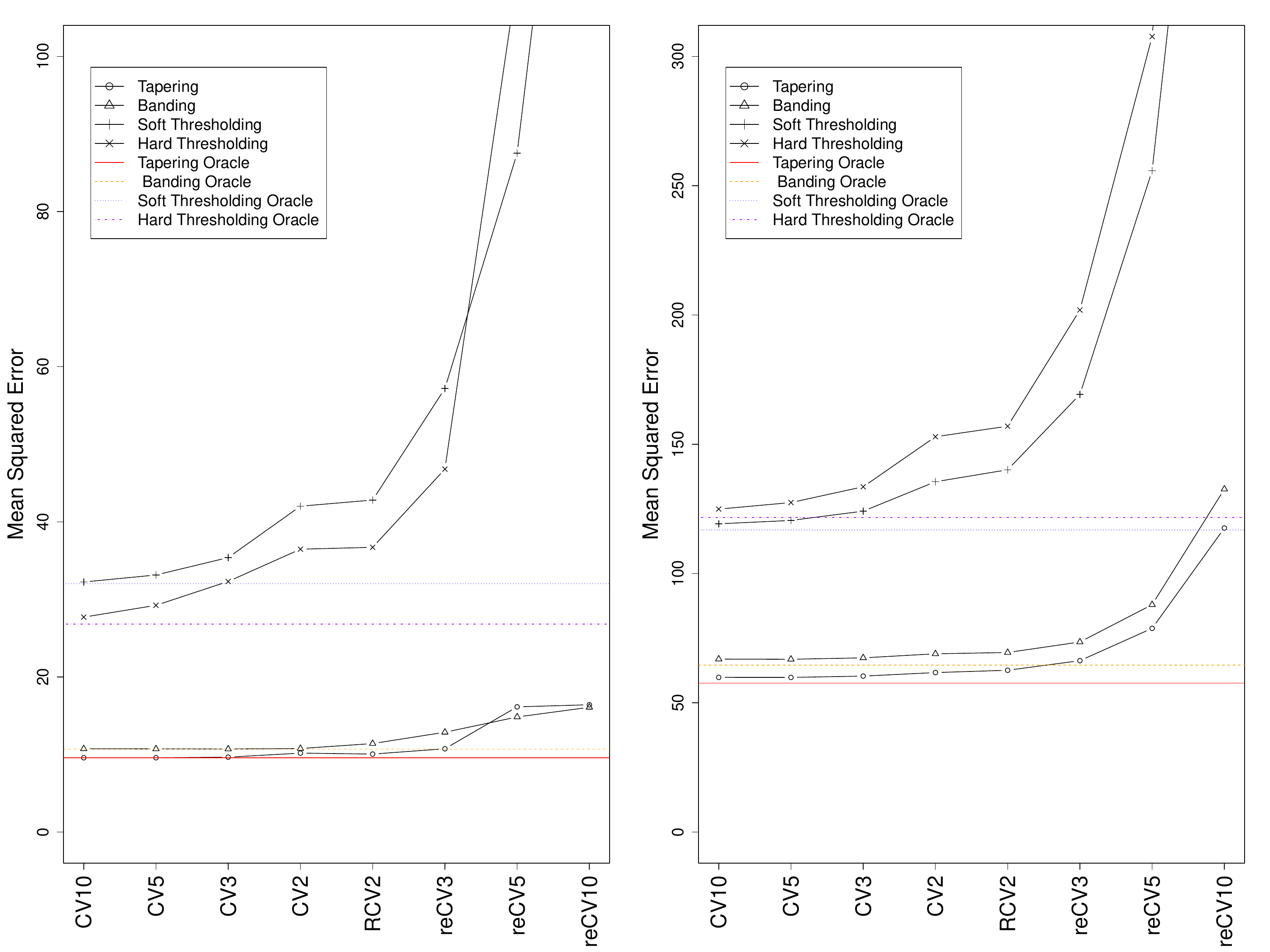}
\end{center}
\end{figure}

\newpage
\begin{figure}[h]\caption{\small Eight cross-validation methods are compared for Model 3 ($n=200$, $p=200$, $\alpha=0.1$ in left panel and $\alpha=0.5$ in right panel). Performances are measured by MSE in Frobenius norm.}\label{F-M3-CV}
\begin{center}
\includegraphics[width=6in, height=4in]{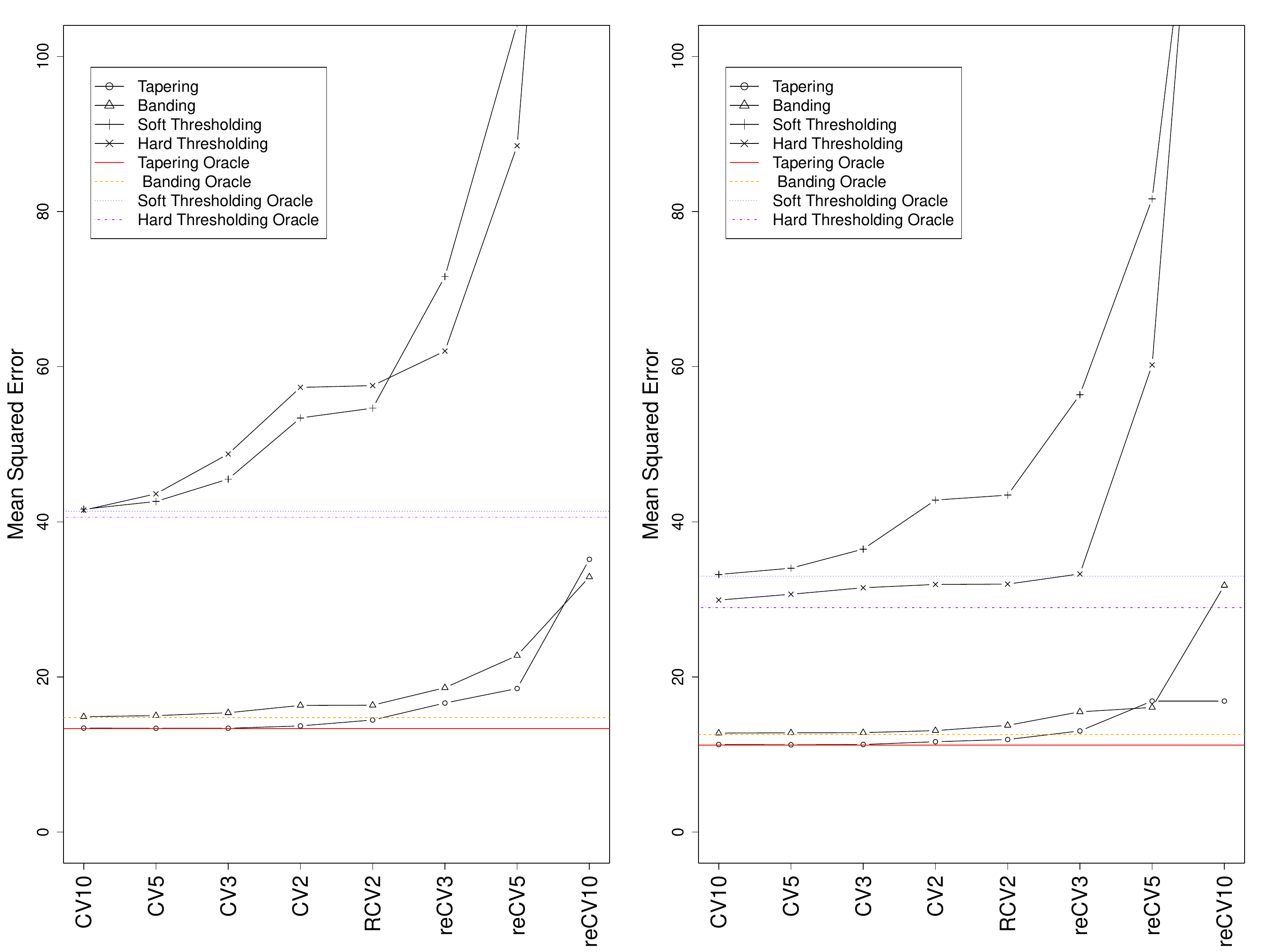}
\end{center}
\end{figure}

\newpage
\begin{figure}[h]\caption{\small  Ten-fold cross-validation is compared with the bootstrap and SURE for Model 1 ($\alpha=0.1$ in left panel and $\alpha=0.5$ in right panel). Performances are measured by MSE in Frobenius norm.}\label{F-M1-boot}
\begin{center}
\includegraphics[width=6in, height=4in]{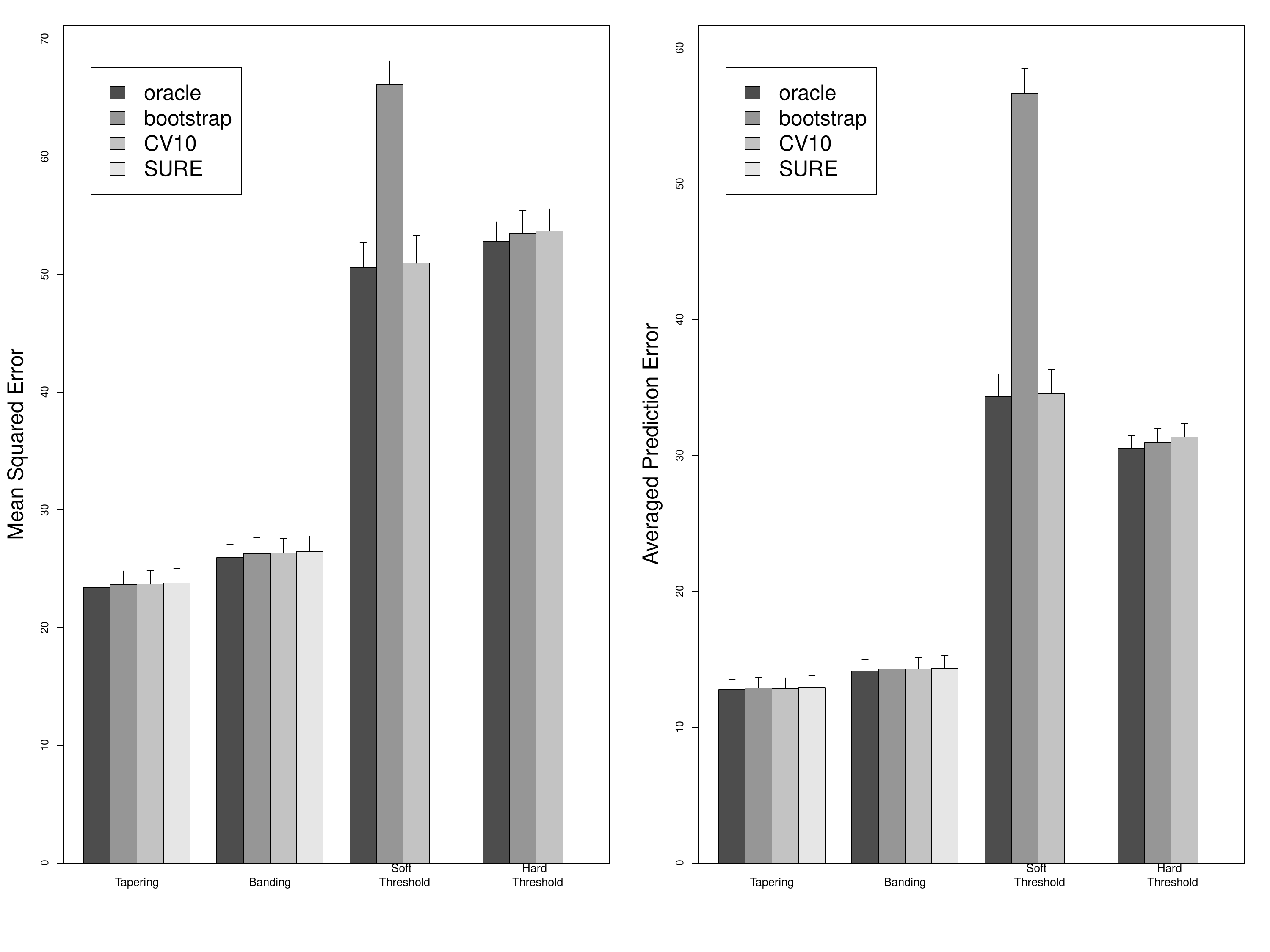}
\end{center}
\end{figure}

\newpage
\begin{figure}[h]\caption{\small Ten-fold cross-validation is compared with the bootstrap and SURE for Model 2 ($\rho=0.9$ in left panel and $\rho=0.5$ in right panel). Performances are measured by MSE in Frobenius norm.}\label{F-M2-boot}
\begin{center}
\includegraphics[width=6in, height=4in]{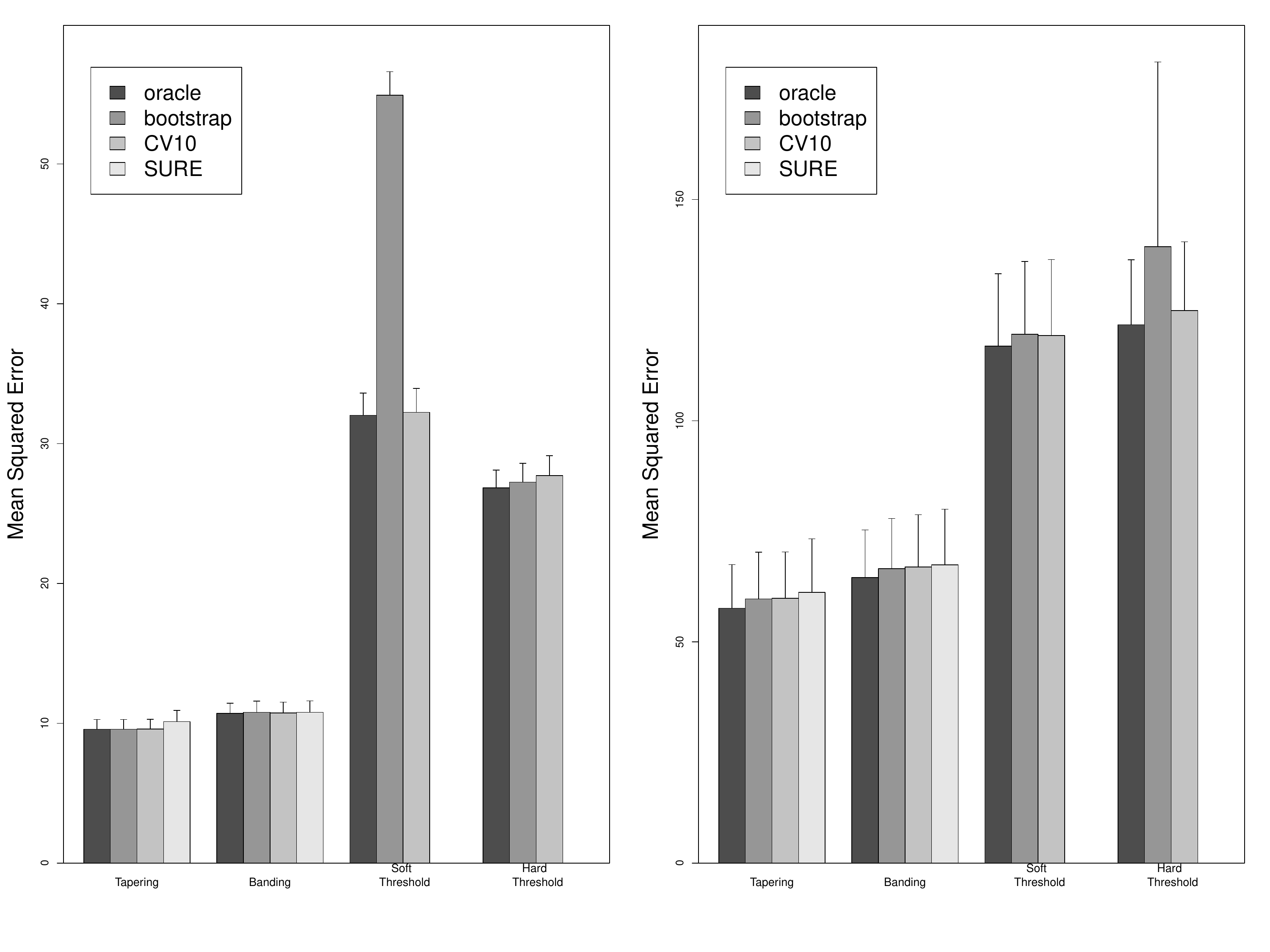}
\end{center}
\end{figure}

\newpage
\begin{figure}[h]\caption{\small Ten-fold cross-validation is compared with the bootstrap and SURE for Model 3 ($\alpha=0.1$ in left panel and $\alpha=0.5$ in right panel). Performances are measured by MSE in Frobenius norm.}\label{F-M3-boot}
\begin{center}
\includegraphics[width=6in, height=4in]{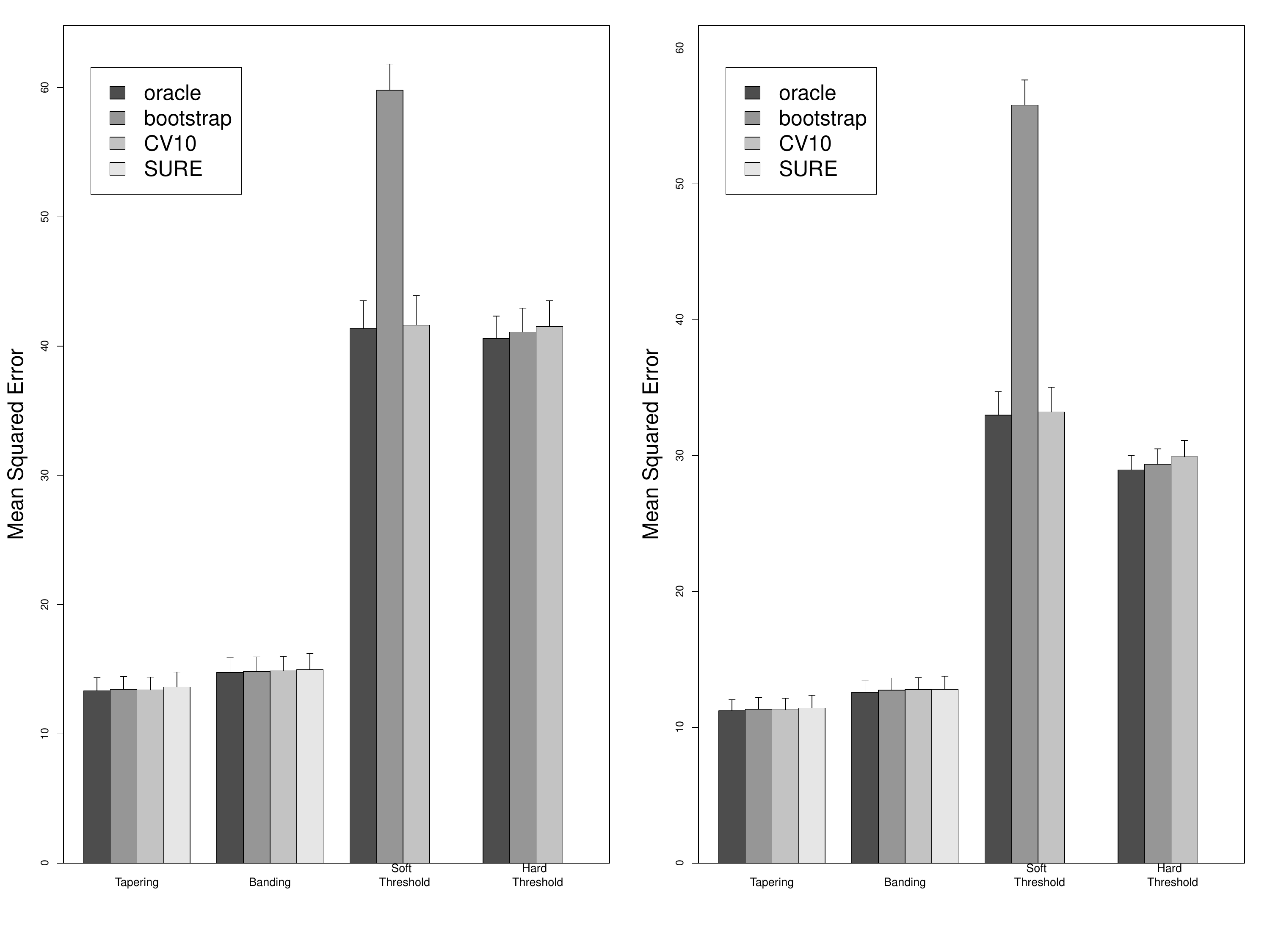}
\end{center}
\end{figure}

\newpage
\begin{figure}[h]\caption{\small Eight cross-validation methods are compared for Model 1 ($n=200$, $p=200$, $\alpha=0.1$ in left panel and $\alpha=0.5$ in right panel). Performances are measured by MSE in operator norm.}\label{L2-M1-CV}
\begin{center}
\includegraphics[width=6in, height=4in]{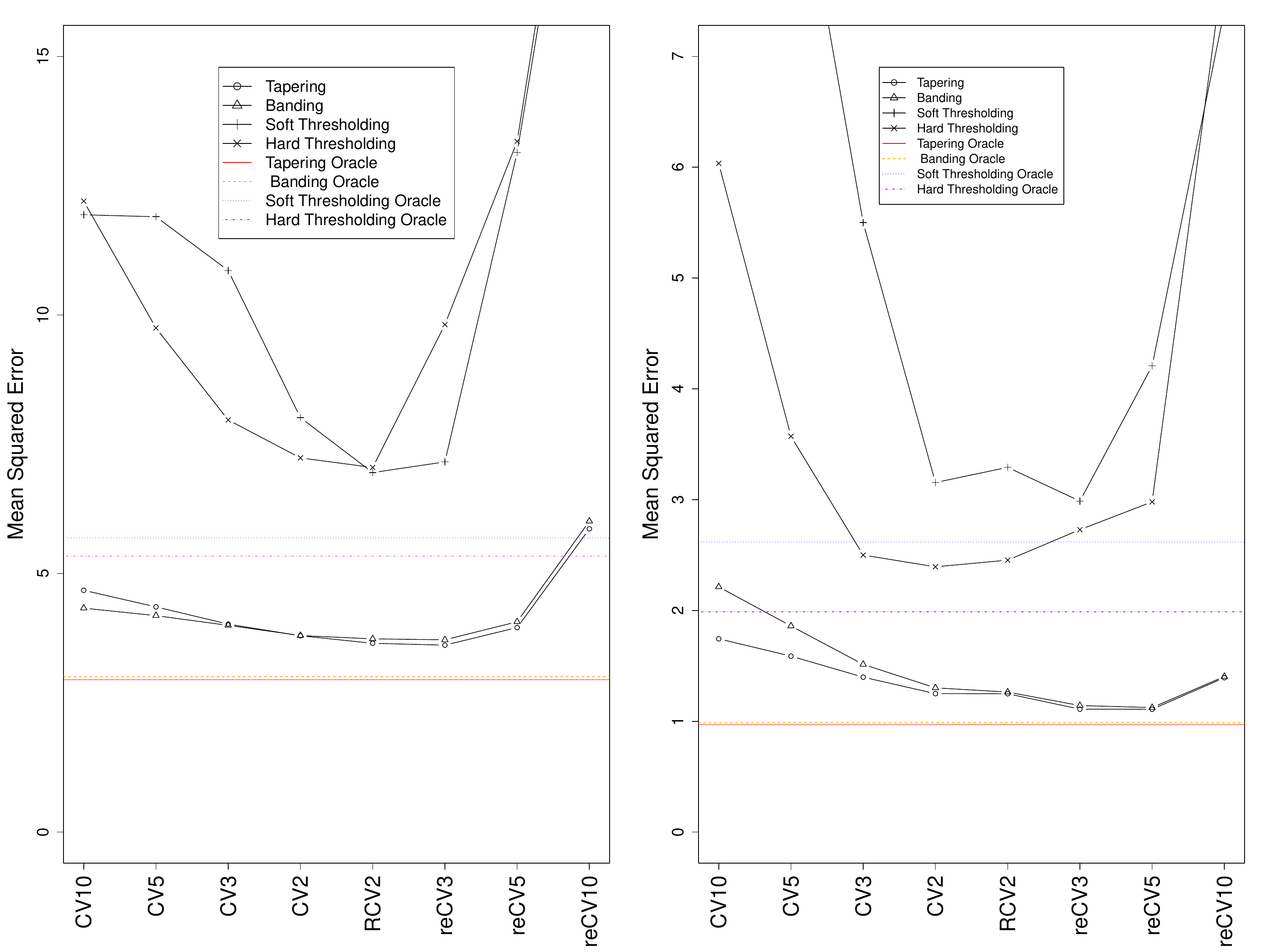}
\end{center}
\end{figure}

\newpage
\begin{figure}[h]\caption{\small Eight cross-validation methods are compared for Model 2 ($n=200$, $p=200$, $\rho=0.9$ in left panel and $\rho=0.5$ in right panel). Performances are measured by MSE in operator norm.}\label{L2-M2-CV}
\begin{center}
\includegraphics[width=6in, height=4in]{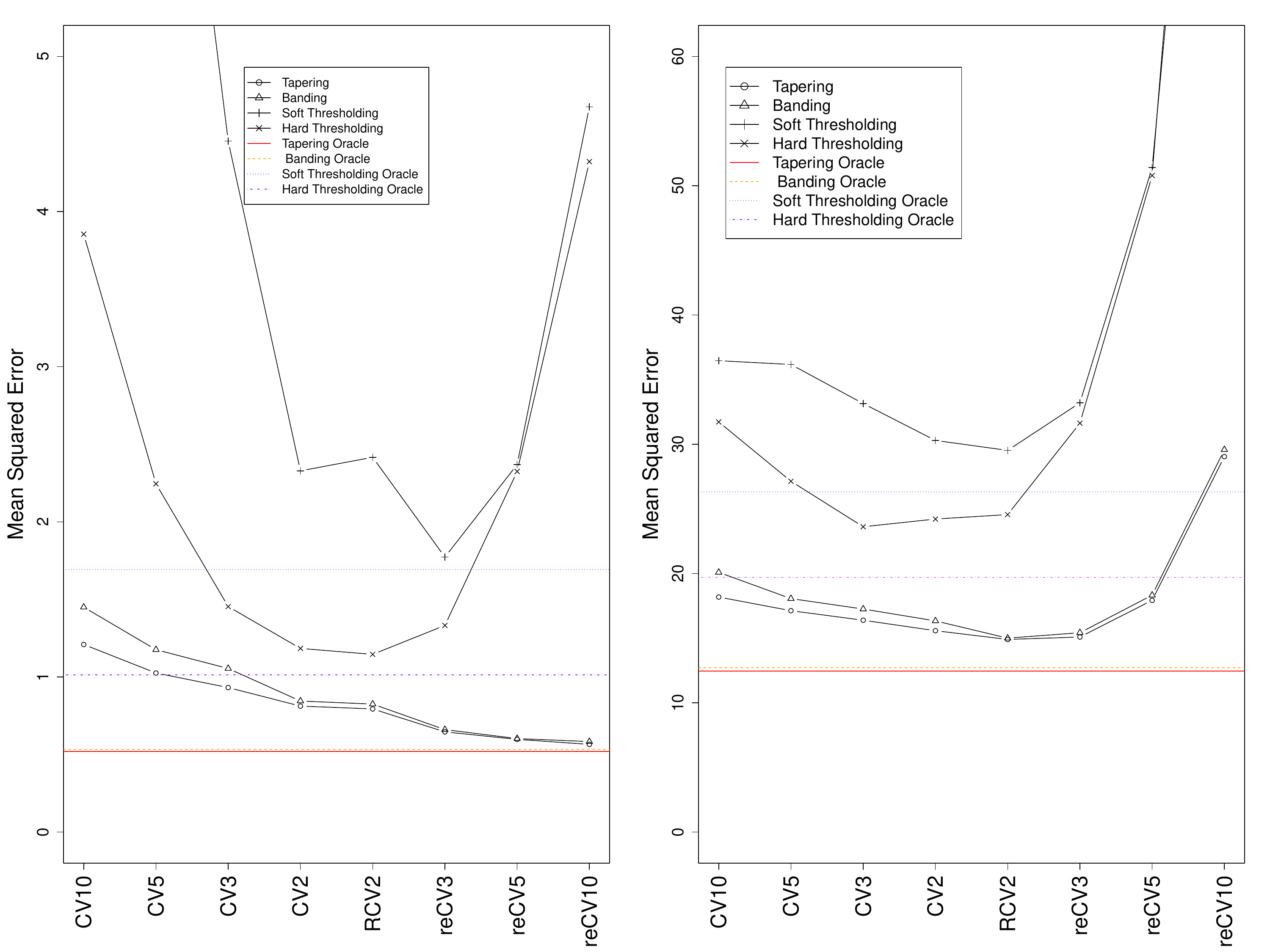}
\end{center}
\end{figure}

\newpage
\begin{figure}[h]\caption{\small Eight cross-validation methods are compared for Model 3 ($n=200$, $p=200$, $\alpha=0.1$ in left panel and $\alpha=0.5$ in right panel). Performances are measured by MSE in operator norm.}\label{L2-M3-CV}
\begin{center}
\includegraphics[width=6in, height=4in]{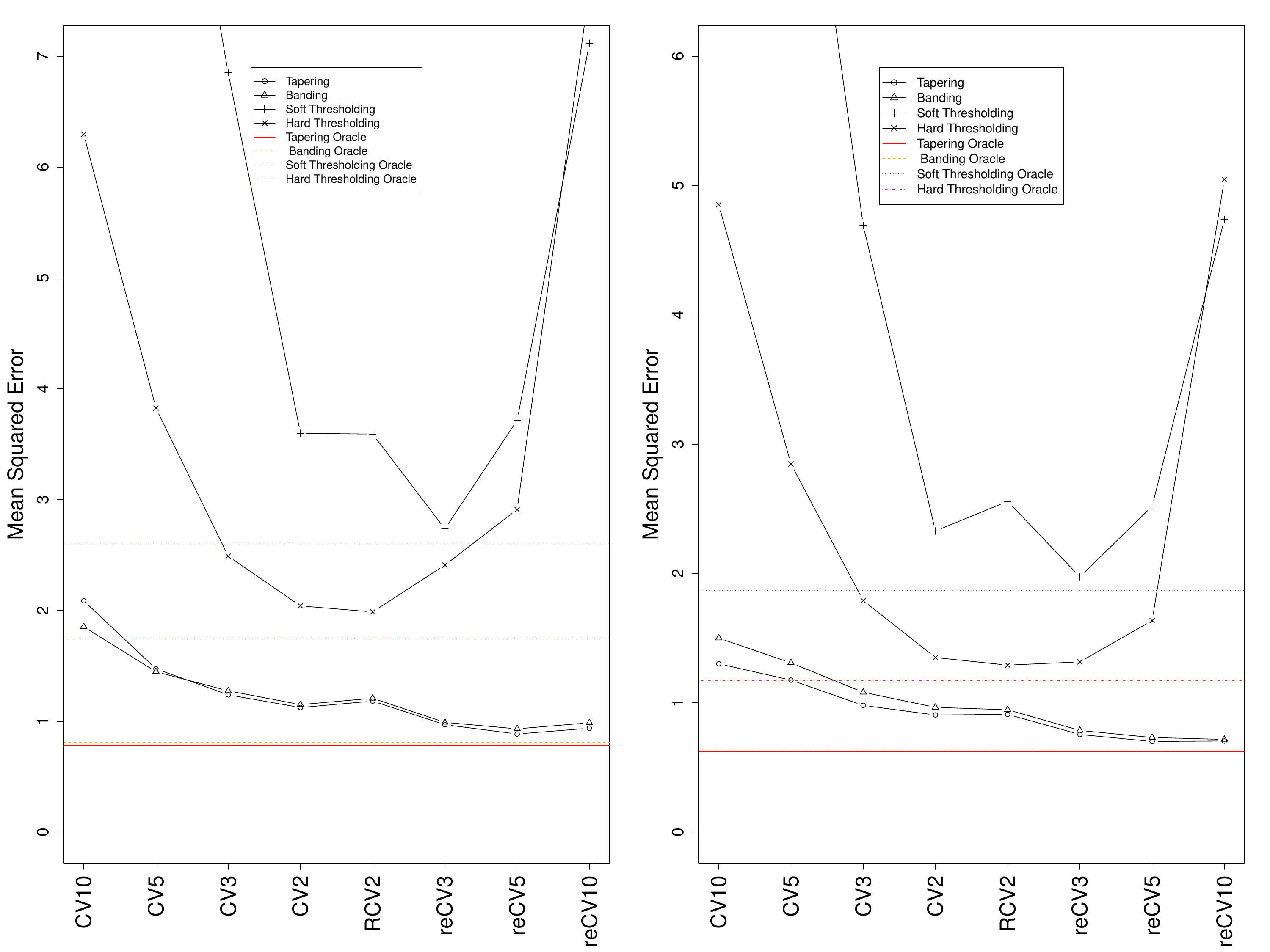}
\end{center}
\end{figure}

\newpage
\begin{figure}[h]\caption{\small  Reverse three-fold cross-validation is compared with the bootstrap method for Model 1 ($\alpha=0.1$ in left panel and $\alpha=0.5$ in right panel). In each panel, the first two bar-graphs are from setting where $n$=100 and $p$=200, while the last two bar-graphs are from setting where $n$=200 and $p$=200. Performances are measured by MSE in operator norm.}\label{L2-M1-other}
\begin{center}
\includegraphics[width=6in, height=4in]{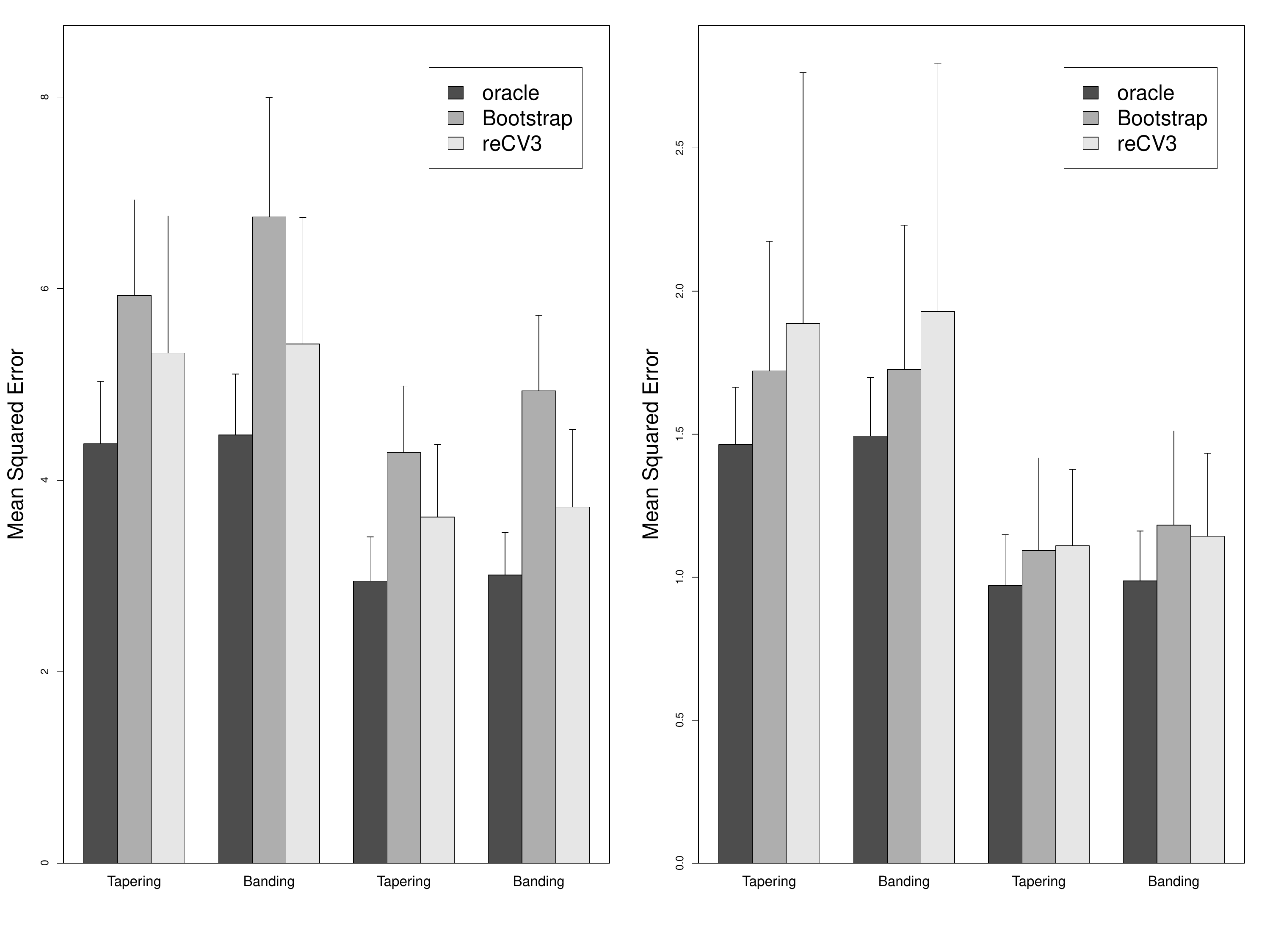}
\end{center}
\end{figure}

\newpage
\begin{figure}[h]\caption{\small  Reverse three-fold cross-validation is compared with the bootstrap method for Model 2 ($\rho=0.9$ in left panel and $\rho=0.5$ in right panel). In each panel, the first two bar-graphs are from setting where $n$=100 and $p$=200, while the last two bar-graphs are from setting where $n$=200 and $p$=200. Performances are measured by MSE in operator norm.}\label{L2-M2-other}
\begin{center}
\includegraphics[width=6in, height=4in]{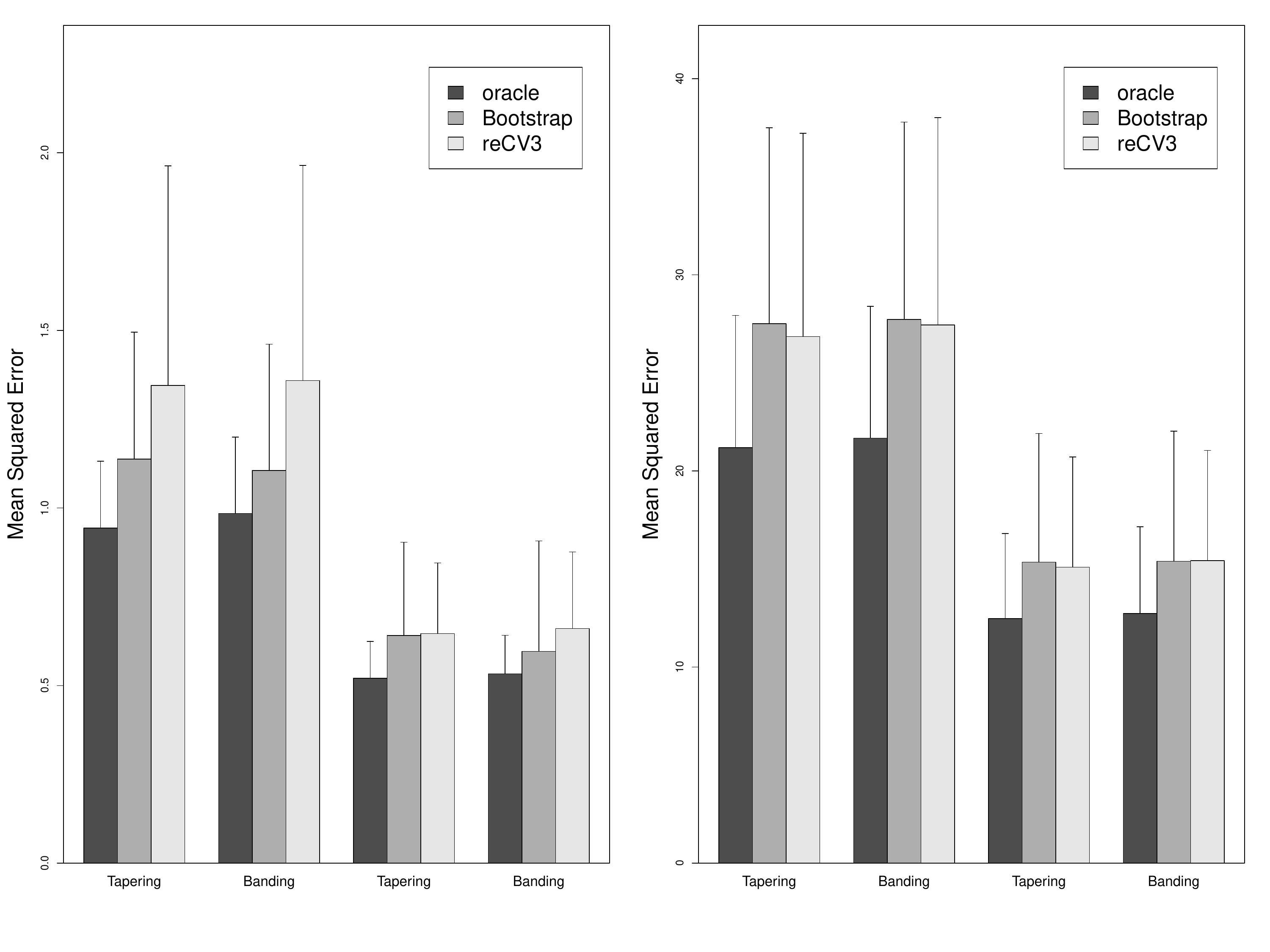}
\end{center}
\end{figure}

\newpage
\begin{figure}[h]\caption{\small Reverse three-fold cross-validation is compared with the bootstrap method for Model 3 ($\alpha=0.1$ in left panel and $\alpha=0.5$ in right panel). In each panel, the first two bar-graphs are from setting where $n$=100 and $p$=200, while the last two bar-graphs are from setting where $n$=200 and $p$=200. Performances are measured by MSE in operator norm.}\label{L2-M3-other}
\begin{center}
\includegraphics[width=6in, height=4in]{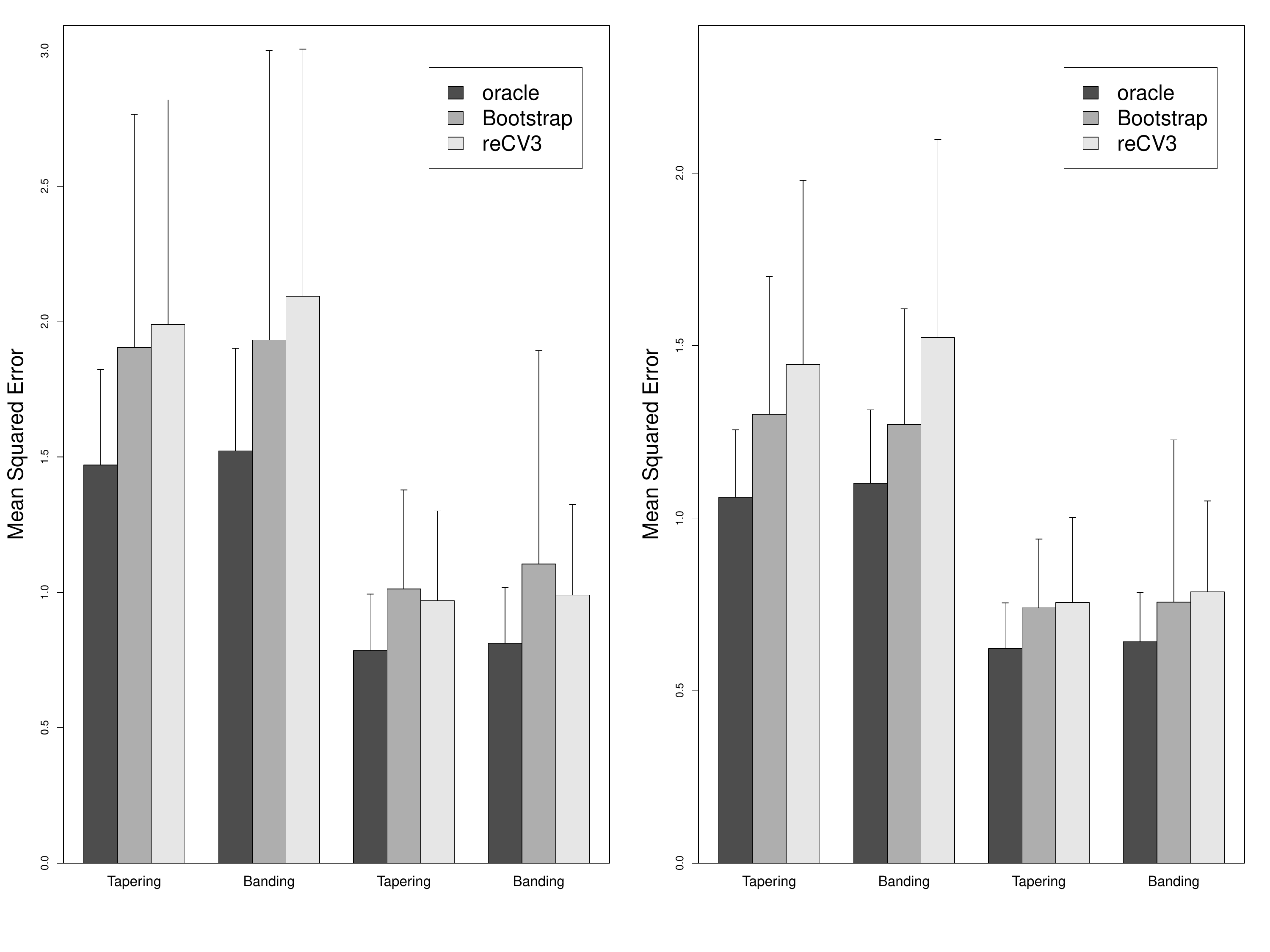}
\end{center}
\end{figure}

\end{document}